# Digital Health and Indoor Air Quality: An IoT-Driven Human-Centred Visualisation Platform for Behavioural Change and Technology Acceptance

Rameez Raja Kureshi
*School of Computer Science,*
*University of Hull,*
*Kingston upon Hull, HU6 7RX, UK*
r.kureshi@hull.ac.uk

Bhupesh Kumar Mishra
*Data Science AI and Modelling Centre*
*(DAIM), University of Hull*
*Kingston upon Hull, HU6 7RX, UK*
bhupesh.mishra@hull.ac.uk

Dhavalkumar Thakker
*School of Computer Science,*
*University of Hull,*
*Kingston upon Hull, HU6 7RX, UK*
d.thakker@hull.ac.uk

Suvodeep Mazumdar
*Information School,*
*The University of Sheffield,*
*Sheffield, UK*
s.mazumdar@sheffield.ac.uk

Xiao Li
*Beijing Capital Online Technology Ltd,*
*Beijing, China*
xiao.li@capitalonline.net

*Abstract*— The detrimental effects of air pollutants on human health have prompted increasing concerns regarding indoor air quality (IAQ). The emergence of digital health interventions and citizen science initiatives has provided new avenues for raising awareness, improving IAQ and promoting behavioural changes. The Technology Acceptance Model (TAM) offers a theoretical framework to understand user acceptance and adoption of IAQ technology. This paper presents a case study using the COM-B model and Internet of Things (IoT) technology to design a human-centred digital visualisation platform, leading to behavioural changes and improved IAQ. The study also investigates users' acceptance and adoption of the technology, focusing on their experiences, expectations, and the impact on IAQ. Integrating IAQ sensing, digital health-related interventions, citizen science, and the TAM model offers opportunities to address IAQ challenges, enhance public health, and foster sustainable indoor environments. The analytical results show that factors such as human behaviour, indoor activities, and awareness play crucial roles in shaping IAQ.

## I. INTRODUCTION

Studies have identified a range of air pollutants, including carbon monoxide (CO), carbon dioxide ($CO_2$), particulate matter ($PM_{2.5}$ and $PM_{10}$), nitrogen dioxide ($NO_2$), sulphur oxides ($SO_x$), and unburned hydrocarbons, as contributors to unhealthy breathing environments [1, 2]. Exposure to these pollutants has been linked to several respiratory issues, such as asthma, cardiovascular problems, and even cancer [3]. Efforts to address air pollution have traditionally focused on outdoor environments. However, research has shown that IAQ can significantly impact human health due to the prolonged exposure individuals experience indoors [4] since people spend the utmost of their time indoors, especially in residential, workplace, and educational settings [5]. Indoor activities such as cooking, cleaning, and ventilation practices can significantly influence IAQ [6]. Additionally, individual behaviours, attitudes, socioeconomic status, and education level are crucial in determining IAQ awareness and maintenance [6, 7]. Because of these, IAQ research has gained momentum in recent years [8, 9], acknowledging the importance of understanding the factors influencing IAQ and developing strategies to improve it. One key strategy is to understand the role of human behaviour and activities in shaping IAQ.

The rise of digital health and behaviour-changing interventions has opened new pathways to enhance indoor air quality and awareness of its impact on human health [10, 11]. These interventions leverage digital platforms to deliver personalised health messages, provide access to specific data, and facilitate behavioural change towards healthier practices. However, the effectiveness of these interventions depends on their ability to citizen engagement and influence their behaviours. In this context, the role of citizens' behavioural science in environmental monitoring and awareness-raising has gained prominence [12, 13] as the citizens' behavioural science initiatives empower individuals and communities to participate in scientific IAQ monitoring research and raising awareness on it, often through the use of low-cost sensor (LCS) technologies and digital platforms [14]. However, the decision to use certain technologies is influenced by their perceived usefulness and ease of use, as well as cost, availability, compatibility with existing systems, and technical support. These factors must be considered when deciding whether to adopt new technologies. The Technology Acceptance Model (TAM) provides a theoretical framework for understanding these factors and predicting user acceptance of technology [15]. This model has been widely applied in various research contexts, including IAQ monitoring and citizen science projects, providing valuable insights into the factors driving user acceptance and technology adoption [15]. The t-tests to analyse participants' evolving acceptance of a digital visualisation platform for household IAQ management over time. T-tests revealed significant increases in Perceived Usefulness (PU: t=-4.90, p=0.00037), Perceived Ease of Use (PEU: t=-2.83, p=0.0152), Attitude Towards Use (AT: t=-6.97, p=0.0000149), and Behavioural Intention (BI: t=-3.24, p=0.00708), indicating enhanced user perceptions and intentions towards the technology over time supported by the feedback got during the semi-structure interviews of the participants. This paper analyses the growing focus on integrating IoT, digital health interventions, citizen science, and the TAM as a user-centric case study for behavioural change in urban living. By integrating these areas, there are opportunities to enhance IAQ, raise awareness, and promote behavioural changes. The analytical insights gained from this integration can provide strategic guidance to researchers and policymakers as they tackle IAQ challenges, improve public health, and create a more sustainable indoor environment. Achieving these goals is possible by utilizing cost-effective IoT technologies and digital platforms and encouraging citizen engagement.

A study was conducted in Bradford involving seven volunteer households to investigate factors that increase awareness of IAQ. The COM-B model has created a human-centred digital visualization platform incorporating IoT technology [10]. This platform provided real-time IAQ-monitored data, information, and advice and was a digital intervention to improve IAQ. A daily digital diary was also introduced to understand participants' indoor activities better. The analytical outcomes show that the digital interventions significantly impacted participants' behaviour, resulting in modifications in indoor activities and overall IAQ improvements. Additionally, we investigated users' acceptance and adoption of the human-centred digital visualisation platforms for IAQ management, focusing on their experiences, expectations, and the technology's influence on IAQ. We employed the TAM to analyse user acceptance for the digital interventions towards measuring IAQ using LCS-based IoT technology.

The rest of the paper is structured as follows: Section II provides a comprehensive review that examines advancements, challenges, and holistic approaches in IAQ, including technology, health impact, data analysis, citizen science participation, and the role of the TAM in understanding user acceptance and adoption of IAQ technology. Then, in Section III, the study design and methodology are presented. Section IV presents the data analysis, experimental results, data interpretation, and study analysis. Section V describes the investigation of users' acceptance and adoption of human-centred digital visualisation platforms for IAQ management, focusing on their experiences, expectations, and the impact of the technology on improving IAQ. Finally, Section VI concludes the paper with a discussion and future work.

## II. LITERATURE REVIEW

### A. Advancements and Challenges in Indoor Air Quality: A Holistic Review of Technology, Health Impact, and Data Analysis

The field of IAQ has been the subject of extensive research, with a significant focus on developing and applying innovative technologies for monitoring and controlling IAQ. A considerable body of literature has explored the use of LCS devices, sensor-based networks, and IoT applications [16, 17]. These studies have highlighted the potential of these technologies in providing real-time data, thereby enabling the effective modelling and management of IAQ. However, the reliability of the data generated by these devices has been questioned. Inconsistencies in calibration and validation methods have emerged as a notable concern, highlighting a critical area for further research [18]. The need for enhanced calibration, validation, and standardisation of sensor performance is a recurring theme in the literature, underscoring the necessity for rigorous methodological approaches in IAQ studies. A subset of the literature has focused on investigating IAQ in specific environments, such as healthcare facilities, university residences, and residential houses [19]. These studies have provided valuable insights into the impact of indoor activities, like cooking and the use of humidifiers, on the generation of indoor particulate matter. Notably, some studies have reported daily average concentrations that exceed the maximum exposure limit recommended by the governments [20]. This highlights the potential health risks associated with poor IAQ and underscores the importance of effective IAQ management. Furthermore, a recurring theme in the literature is the frequent reporting of non-compliance with World Health Organization (WHO) guidelines for IAQ in healthcare facilities [19, 21]. This finding points to a pressing need for stricter adherence to these guidelines in such settings and for policies that promote compliance. The impact of IAQ on human health and cognition has also been a subject of considerable interest in the literature. Studies [22] have revealed that poor Indoor Environmental Quality (IEQ) conditions do not invariably lead to reduced cognition, suggesting that the effects of specific IEQ factors on different cognitive functions are quite distinct. This nuanced understanding of the relationship between IEQ and cognition underscores the complexity of the IAQ-cognition nexus and points to the need for further research in this area. Ventilation is crucial for maintaining good indoor air quality by removing indoor air pollutants and introducing fresh air [23]. Creating a comfortable indoor environment involves human behaviour and architectural design [11]. The factors influencing this process include thermal comfort, ventilation rate, lighting control, and the house's layout. The indoor environment and the IAQ are closely linked and influenced by these specific factors [23]. There are various strategies that experts have implemented to tackle indoor air pollution and improve IAQ. Out of all these strategies, air ventilation has been considered a significant area of focus. Presently, research and governmental initiatives are geared towards enhancing IAQ and fostering behavioural changes. [10]. In order to improve indoor air quality, simple measures like opening windows, using air purifiers, and maintaining ventilation systems can be taken. However, the effectiveness of these measures depends heavily on how well they are communicated and implemented [24, 25]. To ensure the success of a strategy, it is imperative to connect with the intended audience. Without proper engagement, an awareness campaign may fall short of its goals. This is especially true when it comes to IAQ and it is not enough for people to understand the facts; they must also comprehend how IAQ directly affects their health and well-being throughout their lifetime [26].

### B. Exploring Motivations and Approaches in Citizen Science for Environmental Monitoring and Participation

Individuals' involvement in research activities can vary significantly, exhibiting various behavioural traits. Such traits often indicate the extent to which one participates in a research project, ranging from active contributors to less engaged participants and those who solely consume content without contributing. A study [27] reveals that individuals who take part in citizen science can be categorized as Community Workers, Casual Workers or Focused Workers, depending on their participation patterns. While the study provides valuable insights, delving deeper into the factors that drive these participation patterns would have been more beneficial. Since the definitions of participation may differ, it's crucial to comprehend how participants perceive citizen science and the reasons behind their decisions. These motivational factors are intricately linked to participants' emotional, behavioural, cognitive, and social encounters. In another study [28], the authors examined the participation and behavioural characteristics levels in citizen science projects. The study employed a comprehensive methodology, including a literature review, open coding, and the constant comparative method, to identify noteworthy themes from interviews. Robinson et al. [29] introduce a novel approach to user-centred design (UCD) for conveying results in personal exposure studies. Their approach combines human-centred design, human-information interaction, and design thinking to improve participant understanding and engagement. However, it is essential to note that the study's findings are limited to participants in Ljubljana, Slovenia. The study conducted by Golumbic et al. [30] investigates the creation of an air-quality data platform with a user-centric approach. Through three phases of experimentation, the study found that users preferred to interpret data through maps and required data to be contextualized. Despite low registration rates, many users found the platform helpful, indicating that user feedback is crucial in the iterative design of citizen science projects. Hubbell et al. [31] explored the social science aspect of air quality sensors by analyzing people's perceptions, attitudes, and behaviour towards them. They highlighted the potential collaboration between citizen scientists and professionals to improve the understanding of sensor technology usage and increase public awareness of air quality issues. Pritchard et al. [32] analyze how citizen sensing can address issues like air pollution, developing new technologies, partnerships, and communities. Another study conducted by Mahajan et al. [33] studied designing a citizen-driven framework for monitoring air quality in Taiwan. The study emphasizes the importance of collective awareness and knowledge sharing and highlights the transition from a Do-It-Yourself (DIY) approach to a Do-It-Together (DIT) approach. One of the case studies explored in the document concerns the practice of incense burning in religious contexts. The study utilized AirBox devices to examine and discuss the link between cultural practices and science. It stresses the significance of understanding how cultural practices impact air quality and how citizen science can help to increase awareness about this issue. English et al. [34] highlight the importance of involving the community in all aspects of air quality monitoring when designing a community-wide monitoring program. This emphasizes the need for collaborative approaches in environmental monitoring. Finally, Leonardi et al. [35] propose a mobile crowdsensing system for air quality monitoring, which enables participants to reflect on their exposure to pollution. This highlights the potential of crowdsourcing in collecting valuable data and insights on air pollution.

### C. The Technology Acceptance Model: A Lens into User Acceptance and Adoption of Technology

The TAM model is a valuable tool for anticipating user reactions to information systems. It posits that users' decisions to accept or reject a new technology hinge primarily on two factors: the perceived usefulness and perceived ease of use. In essence, if users view a new technology as helpful and user-friendly, they are more inclined to embrace it. [15]. Over time, the TAM has been expanded and adjusted to cover more aspects that can affect technology acceptance. Even though it is a simple model, TAM has been extensively applied in various research settings and has proven reliable in anticipating user acceptance of technology. In the review of TAM by Yousafzai et al. [36], a unified model was proposed that integrates TAM with other user acceptance models, providing a complete understanding of technology acceptance. For example, Zhou [37] used the TAM to study mobile payment adoption. The study included factors like perceived credibility and personal innovativeness in the field of Information Technology (IT). The results showed that the model could be applied to different technology contexts, which is useful for understanding how people use technology in their daily lives. Laukkanen et al. [38] found that people's resistance to using new technology is often overlooked. They discovered that concerns about potential problems, being clear about what their role is when using the technology, and whether they feel confident in their ability to use it are all important factors that influence people's willingness to accept new technology. This study helps us better understand why some people might be hesitant to use new technology. Kim et al. [39] emphasize user autonomy in technology acceptance by introducing the concept of perceived discretionary power of users into TAM. Taherdoost [40] discusses several theories and models regarding technology acceptance, including TAM. This research highlights the significance of TAM in the area of technology adoption and emphasizes the need to comprehend the factors that influence users' acceptance or refusal of technologies. Further findings have applied TAM in various contexts, such as mobile health interventions in resource-limited settings [41], Fintech services [42], e-commerce/e-business technology among small and medium enterprises [43], and educational technology [44]. Venkatesh et al. [45] propose a theoretical extension of TAM that elucidates the influence of social factors and cognitive processes on perceived usefulness and usage intentions. Al-Rahmi et al. [46] delve into the behavioural intentions of students to utilize social media in higher education, as well as their actual usage. Similarly, the concept has been applied in the healthcare sector to comprehend the uptake of various platforms such as Electronic Health Records [47], telemedicine [48], mHealth apps [49], patient portals [50], wearables [51], and AI-driven tools [52].

To summarize, the papers provide a detailed and comprehensive understanding of the TAM and its extensions. They highlight the significance of multiple factors that play a crucial role in comprehending user acceptance of technology, such as perceived ease of use, usefulness, role clarity, risk, self-efficacy, personal innovativeness, and discretionary power. These findings can assist in developing and implementing technology that ensures higher user acceptance.

### III. METHODOLOGY

This paper aims to evaluate the effectiveness of an IoT device in monitoring IAQ. Additionally, it examines the role of a human-centred digital visualization platform in raising participant awareness levels through digital interventions. Also, to investigate how these technologies impacted participants' behaviour to improve IAQ. This study obtained ethical approval from the University's Research Ethics Panel for the Biomedical, Natural, Physical, and Health Sciences.

*A. Study Design*

Our previous work allowed us to develop LCS-based IoT devices that reliably monitor IAQ [10]. These devices can monitor air pollutants, including particulate matter ($PM_{2.5}$ & $PM_{10}$), and meteorological parameters, such as temperature and relative humidity. Using machine learning techniques, these devices are calibrated against high-fidelity reference AQ monitoring stations to measure particulate matter. These calibrated devices require a main power supply and Wi-Fi for data transmission. They are programmed to collect and transmit IAQ data to a cloud server for processing every 15 minutes.

The COM-B model and Behaviour Change Wheel (BCW) framework from our prior research were used to develop digital interventions as an interactive web page to augment user experience and foster an elevated level of awareness about IAQ among the study participants [10]. These visualise IAQ-monitored data and record everyday indoor activities, as shown in Figure 1-5. The customized platform designed for this study was instrumental in providing users with a more comprehensive understanding of IAQ dynamics. The visualisation was built to present PM data with five plots covering the WHO limit, the UK limit, today's average value, the week's average value, and the previous week's average value. The set of five bar graphs provided enables participants to assess and compare their indoor air pollution levels with two standard guidelines from the WHO and the UK. This allows for a more detailed interpretation of the data, which in turn provides a better understanding of the potential health risks associated with the levels of indoor air pollution. In addition, this digital diary includes interactive multiple-choice questions in three steps regarding indoor activities such as vacuum cleaning, opening windows, smoking, breathing problems, cooking and heating, as presented in Figure 5. These questions are delineated based on the literature on air quality-related health impacts and socio-diversity [53, 54].

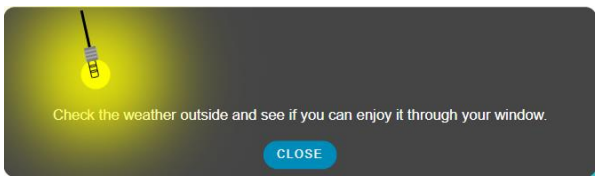

Fig. 1. A pop-up message whenever the participant login to the platform

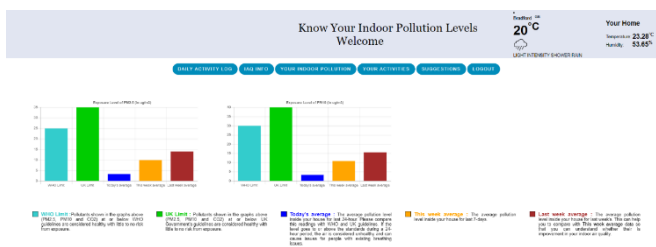

Fig. 2. Human-centred digital visualisation platform

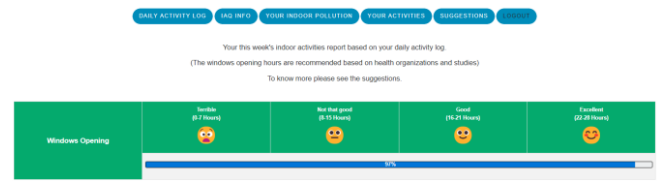

Fig. 3. Indoor activity – window control for ventilation

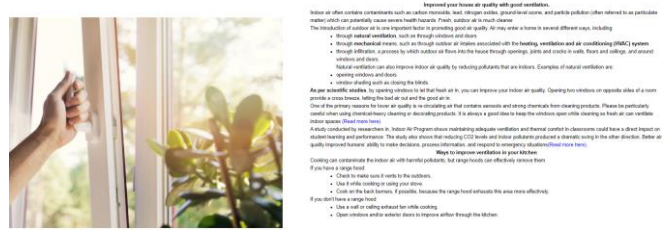

Fig. 4. Provide IAQ-related information, including improvement and health impacts.

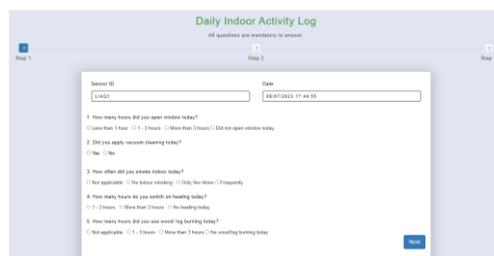

Fig. 5. The daily digital diary includes multiple-choice questions related to indoor activities.

*B. Study Setting, Participants, and Context*

As part of the SCORE and LifeCritical projects funded by the European Union, research was conducted in Bradford, a city located in the north of the UK. The community group was selected with the help of the Bradford Metropolitan District Council (BMDC), which is a partner in both projects and assisted in the initial contact between the researchers and the participants.

The study was conducted in June 2023 over a period of three weeks, as shown in Figure 6. The primary purpose was to observe changes in participant behaviour when they were provided with digital interventions based on the COM-B model to improve indoor air quality. The focus was on identifying the digital intervention that had the most significant impact on changing their behaviour. Also, to measure the acceptance of this technology, including human-centred digital visualisation platforms. In-person workshops were held to provide an overview of the study, including information about the device and the deployment process. The workshops stimulated significant participant engagement, as evidenced by the numerous questions posed by participants concerning indoor air quality and its significance, the benefits of monitoring IAQ, and related topics. This interest led to a willingness among citizens to participate in the study. In order to conduct the study, the research team installed LCS-based IAQ monitoring devices in seven selected households. The selection criteria encompassed a variety of socioeconomic and demographic factors, including location, ethnicity, and type of dwelling. During the first week of the study, participants were not provided access to their IAQ data, nor were they required to keep a daily digital diary of their indoor activities. In the second week, participants were given access to their IAQ data and introduced to the digital visualization platform. Following the third week, the research team scheduled online meetings with participants at convenient times to conduct interviews using pre-prepared questions. After the third week, the IAQ monitoring devices were collected from the participants' homes. The team remained available and maintained regular contact with participants to provide technical assistance for accessing the platform or completing the diaries. Participant IDs were anonymised to prevent linkage with specific individuals.

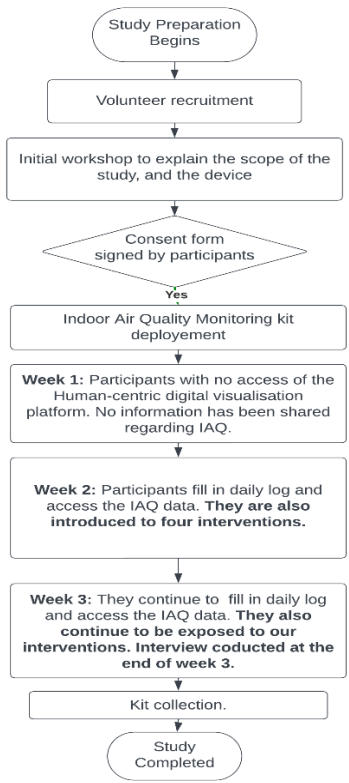

Fig. 6. Diagram illustrating the flow of the study

## IV. DATA ANALYSIS AND DISCUSSION

After precisely gathering and analyzing IAQ and indoor activity data from participants' homes through our digital visualization platform for three weeks, we have gained valuable insights and information.

### A. Analysis of Improvement in IAQ Based on Behavioural Change

During a three-week study, IAQ data was collected and analyzed using a human-centred digital visualization platform based on the COM-B model and digital interventions introduced in the second week. The dataset included PM2.5 and PM10 measurements from various devices over the three weeks. The IAQ data was thoroughly analyzed using multiple methods to uncover patterns and trends within the dataset. This analysis involved examining the IAQ data over the entire three-week period. These diverse analytical approaches provided a comprehensive understanding of the IAQ data, forming a strong basis for further interpretations and conclusions.

Visual analysis using bar graphs was conducted to show the average $PM_{2.5}$ and $PM_{10}$ levels of each device over three weeks, as seen in Figures 7 and 8. The findings indicated a significant difference in $PM_{2.5}$ and $PM_{10}$ levels in various households (L1-L7), implying that the indoor environment impacted indoor air pollution levels. Additionally, fluctuations in $PM_{2.5}$ and $PM_{10}$ levels were observed in each household weekly, indicating changes in environmental conditions, indoor activity patterns, or other factors that affected indoor air pollution levels. Despite these variations, a consistent pattern was observed across all devices. Specifically, PM levels were higher in the initial week and lower in the subsequent and final weeks, suggesting a common factor influencing indoor air pollution levels across all devices, such as window opening, cooking, cleaning, and other indoor activities.

Throughout the study, implementing a digital visualization platform and digital interventions beginning in the second week resulted in a notable decrease in indoor air pollution levels. This decrease persisted into the third week, indicating that the impact of the digital visualization platform and interventions was more than temporary but instead had a lasting effect. To support the visual analysis, independent two-sample t-tests were employed to compare the means of $PM_{2.5}$ levels between different weeks. The results revealed a statistically significant difference in the means of $PM_{2.5}$ levels between Week 1 and Week 2 (t-statistic: 4.20, p-value: 0.000027), Week 2 and Week 3 (t-statistic: 2.85, p-value: 0.0044), and Week 1 and Week 3 (t-statistic: 6.05, p-value: 0.00000000166). As the p-values were less than 0.05, which is commonly used as a threshold for statistical significance, it suggests that the differences in $PM_{2.5}$ levels between the weeks were not due to chance but rather were likely due to the introduction of the digital visualization platform and interventions. Similarly, the results showed a significant difference in the means of $PM_{10}$ levels between Week 1 and Week 2 (t-statistic: 4.17, p-value: 0.000032), Week 2 and Week 3 (t-statistic: 2.81, p-value: 0.0050), and Week 1 and Week 3 (t-statistic: 6.00, p-value: 0.00000000227), providing statistical evidence that the interventions had a significant impact in reducing $PM_{10}$ levels as well. As such, both the visual and statistical analyses suggest that the introduction of the digital visualization platform and digital interventions had a significant impact in significantly reducing indoor air pollution levels.

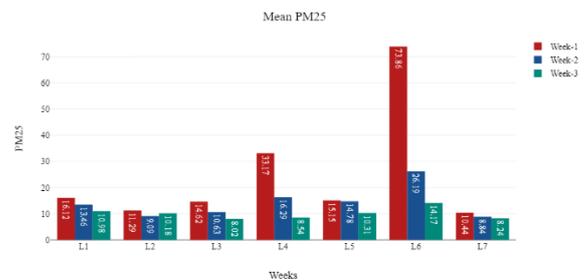

Fig. 7. Mean $PM_{2.5}$ data from all households over three sequential weeks.

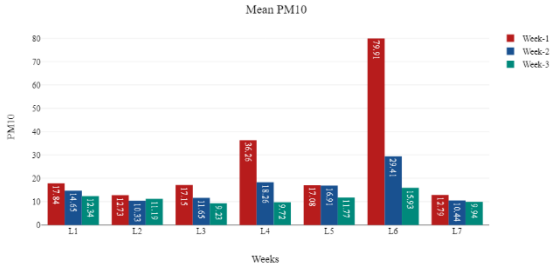

Fig. 8. Mean PM$_{10}$ data from all households over three sequential weeks.

## V. ROLE OF THE TECHNOLOGY: EXPERIENCE & EXPECTATIONS

Another aim of this research study is to investigate users' acceptance of digital visualisation platforms as a means to increase their understanding and management of IAQ in their houses. To achieve this, the study incorporated the TAM [15, 55]. According to the TAM, user acceptance of digital platforms depends on factors such as perceived usefulness (PU), perceived ease of use (PEU), attitude intention (AI), and behavioural intention (BI) towards technology adoption, as illustrated in Figure 9. The TAM provides a theoretical framework for understanding users' attitudes and behaviours towards adopting and utilising new technologies. In the context of this study, perceived usefulness refers to users' perception of how the digital visualisation platform can effectively contribute to improving their understanding and management of their household IAQ. The perceived ease of use pertains to the user's perception of the platform's usability and simplicity.

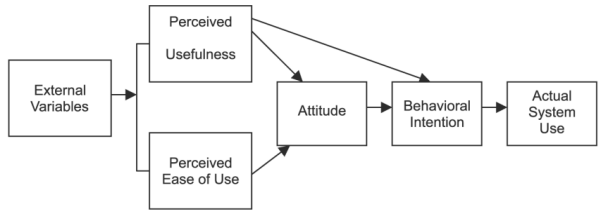

Fig. 9. Diagram of TAM [55]

The TAM provides a theoretical framework to comprehend users' attitudes and behaviours towards adopting and using new technologies. In this study, perceived usefulness refers to users' perception of how effectively the digital visualization platform can improve their understanding and management of their household IAQ. The perceived ease of use affects the user's perception of the platform's simplicity and usability. These factors, perceived usefulness and ease of use are crucial in determining users' attitudes and intentions towards the platform and the technology.

In addition, the attitude intention of users is their overall evaluation and subjective judgment towards a digital visualization platform. This evaluation reflects their attitude and inclination towards adopting and using the technology. On the other hand, behavioural intention represents users' willingness and readiness to engage in specific behaviours related to the technology, such as actively utilizing the platform for IAQ monitoring and making informed decisions based on the provided data. The objective of this study is to gain insights into users' acceptance of a human-centred digital visualization platform. The study incorporates the TAM model and examines the relationships between PU, PEU, AI, and BI. The findings of this study will provide information about the factors that affect users' attitudes, intentions, and behaviours towards adopting and utilizing the platform to improve their understanding and management of IAQ in their homes.

After the introduction of the digital visualisation platform, the TAM questionnaire (Table 1) was administered to households on two occasions. The first questionnaire was given during the second week when households were exposed to the digital interventions for the first time. The second questionnaire was administered at the conclusion of the study. The purpose of this study was to track any changes in the participants' willingness and interest in using the platform over an extended period of time. By comparing their initial perceptions and expectations with their experiences and attitudes towards the platform at the end of the study, we gained valuable insights into the platform's effectiveness and impact. The repetitive administration of the TAM questionnaire allowed for a thorough analysis of the participants' evolving acceptance and intention, thereby improving the study's understanding of the platform's impact on household indoor air quality management.

TABLE I. TAM-BASED STUDY QUESTIONNAIRE

| *Elements* | *What are we evaluating?* |
|---|---|
| Perceived Usefulness (**PU**) | **PU1:** How useful do you find the indoor air quality data and information provided on our platform? <br> **PU2:** Which indoor air quality improvement suggestion did you find most helpful to you? <br> **PU3:** Did you find our platform's indoor activity tracking feature helpful? <br> **PU4:** How accurate do you find the indoor air quality data provided on our platform? |
| Perceived Ease of Use (**PEU**) | **PEU1:** How easy was it to navigate the indoor air quality platform? <br> **PEU2:** How easy was tracking your indoor activity on our platform (Windows opening)? <br> **PEU3:** Did you encounter any difficulties in using our web platform? |
| Attitude (**AT**) | **AT1:** How positively or negatively do you feel about using our indoor air quality platform regularly in the future? |
| Behavioural Intention (**BI**) | **BI1:** How likely are you to use our indoor air quality platform again tomorrow to improve indoor air based on the information available? |

### A. Quantitative TAM data analysis using a questionnaire-based approach

The following t-tests compare the means of the first and second responses for each component: PU, PEU, AT, and BI.

a) **Perceived Usefulness (PU):** The t-statistic of -4.90 and the p-value less than 0.05 provide evidence of a statistically significant improvement in perceived usefulness between the first and second responses. This indicates that participants found the human-centred digital visualization platform more valuable over time.

b) **Perceived Ease of Use (PEU):** The analysis shows a significant increase in perceived ease of use from the second week of introduction to the end of the research period. The t-statistic value is -2.83, and the p-value is 0.0152, representing that the observed change is not due to random chance.

c) **Attitude Towards Use (AT):** Based on the t-statistic of -6.97 and the p-value of 0.0000149, it is evident that participants' attitudes towards technology have significantly improved over time. This finding emphasizes the significance of adopting technological advancements and highlights the potential advantages of having an open and receptive mindset towards new technologies.

d) **Behavioural Intention (BI):** Based on the analysis, we found that the t-statistic is -3.24, and the p-value is 0.00708. This statistical information indicates that there has been a significant increase in the participants' behavioural intention from the second week to the third week. Hence, the result suggests that there is a notable rise in the participants' intention to use the technology over time.

In summary, the results strongly suggest that with time, the technology's perceived usefulness, ease of use, and the user's intention and attitude to use it all showed significant improvement.

*B. Qualitative TAM analysis using Semi-structured Interviews.*

From the responses provided by the participants, it is clear that the human-centred digital visualization platform used to raise awareness about IAQ was well-received. We conducted semi-structured interviews to explore this technology's acceptance and satisfaction levels. The participants expressed high confidence in the technology and recognized its effectiveness in raising awareness about IAQ at the household level. Our qualitative analysis revealed some key findings:

*1)* **Affirmative Response to Technology:** All participants exhibited a positive response toward using technology for IAQ monitoring at the household level. This sentiment is encapsulated in expressions such as, *"I would say big Yes to this technology"* and *"Well, yes, I do like to use the platform."* These responses underscore a significant level of technology acceptance, resonating with the TAM principle of PU.

*2)* **Increased Awareness and Understanding:** The participants articulated that the platform facilitated a more profound comprehension of their domestic Indoor Air Quality (IAQ) scenario and the avenues for its enhancement. This is reflected in remarks such as, *"This aids significantly in comprehending my household circumstances and the measures that can be adopted for ameliorating the IAQ, courtesy of the platform."* Such feedback intimates that the platform effectively heightened awareness and proffered pragmatic recommendations, aligning well with the study's primary objectives.

*3)* **Confidence in the Technology:** Participants articulated a high level of assurance in the AQ monitoring technology. Statements such as *"To be honest. I like the study, so I can say that I have enough confidence"* and *"Yes, I have confidence with this technology. This platform helps me so much to understand about my house's air quality"* underscore a profound trust in the capabilities of the technology. This sentiment resonates with the TAM principle of PEU.

*4)* **Sharing with Social Circle:** Some participants disseminated the knowledge acquired from the study to their social circles, as illustrated by the following statement: *"I did tell my friends about the device like how it measures the quality of the house and based on that how we can take preventive actions, you know, in order to improve the overall air quality of the house."* This suggests that the platform not only had a profound influence on the participants but also extended its impact to their acquaintances and family members.

*5)* **Behavioural Changes:** A participant shared that they have changed their behaviour regarding vaping after being informed about its adverse health effects through the platform. This demonstrates that the platform has successfully facilitated behavioural changes, which aligns with its primary goal of utilizing the COM-B model and BCW framework supported by IoT technology.

In summary, participant feedback suggests that the IAQ monitoring technology and digital interventions were highly regarded, fulfilling expectations and inspiring trust. The platform effectively raised awareness, delivered practical recommendations, prompted behavioural changes, and yielded beneficial outcomes for users and their communities. These results affirm the efficacy of the human-centric digital visualization platform and its potential for widespread use in IAQ monitoring and enhancement.

## VI. Conclusion & future work

This research has contributed to the field of IAQ awareness by demonstrating the effectiveness of a digital visualisation platform in raising behavioural change using technology in urban living. The platform's user-centric design and features have proven instrumental in fostering active participant engagement. The heightened level of engagement has effectively raised awareness and educated individuals about the significance of IAQ, consequently resulting in the adoption of healthier practices. The platform has played a crucial role in promoting awareness and education about IAQ among its users and their social circles. Participants have admitted sharing information, experiences, and insights on IAQ through the platform, leading to more informed discussions and actions. This has resulted in a ripple effect, with the impact of the platform's efforts reaching far beyond its immediate users. The study has underscored the importance of continuous user feedback in developing and refining such platforms, with participants' suggestions for improvements providing valuable insights for future development in real-world application areas such as monitoring health status, raising awareness on specific citizen issues and other interactive studies. Moreover, the research has shown that digital interventions can help to promote better practices for improving IAQ. The platform has significantly increased user engagement and effectiveness by personalizing the interventions according to each user's unique needs and preferences. This study provides a strong foundation for using digital tools to promote environmental health and sets the stage for further exploration into how these tools can be optimized to create healthier urban living environments. T-tests revealed significant increases in Perceived Usefulness (PU: t=-4.90, p=0.00037), Perceived Ease of Use (PEU: t=-2.83, p=0.0152), Attitude Towards Use (AT: t=-6.97, p=0.0000149), and Behavioral Intention (BI: t=-3.24, p=0.00708), indicating enhanced user perceptions and intentions towards the technology over time supported by semi-structure interview analysis.

It is imperative to consider certain limitations that the research may encounter. Firstly, the sample size is small, consisting of only seven households. Secondly, the long-term accuracy of the LCS used in the study may be affected by the need for re-calibration. Thirdly, the study duration was relatively short, lasting only three weeks. Lastly, the study only measured a limited number of factors in indoor air quality, indicating that more extensive and longer-term studies conducted across multiple seasons are needed to understand the issue better.

Recognising these limitations, future work offers several promising avenues for exploration. A deeper understanding of the influence of IAQ on human health and perception is crucial. The digital visualisation platform can be refined based on participant feedback, with potential features like a reminder system for daily logs, mobile alerts, and broader compatibility. Moreover, expanding the study to include a more diverse range of participants and environments will provide a holistic view of the challenges and opportunities in enhancing IAQ awareness. This research is a step in the ongoing scientific journey to understand and improve indoor air quality through digital interventions.

ACKNOWLEDGEMENT

This research work is funded by the European Commission Interreg project Smart Cities and Open data Reuse (SCORE) and LifeCritical Projects.

REFERENCES

[1] A. I. Mansour and H. A. E. Al-Jameel, "Traffic flow impacts on the environment along urban streets under mixed traffic conditions in Al-Najaf city," in *IOP Conference Series: Earth and Environmental Science*, 2023, vol. 1129, no. 1: IOP Publishing, p. 012046.

[2] J. Gonzalez-Martin, N. J. R. Kraakman, C. Perez, R. Lebrero, and R. Munoz, "A state–of–the-art review on indoor air pollution and strategies for indoor air pollution control," *Chemosphere,* vol. 262, p. 128376, 2021.

[3] S. Steinle *et al.*, "Personal exposure monitoring of PM2.5 in indoor and outdoor microenvironments," *Science of The Total Environment,* vol. 508, pp. 383-394, 2015/03/01/ 2015, doi: https://doi.org/10.1016/j.scitotenv.2014.12.003.

[4] J. Saini, M. Dutta, and G. Marques, "A comprehensive review on indoor air quality monitoring systems for enhanced public health," *Sustainable Environment Research,* vol. 30, no. 1, p. 6, 2020.

[5] M. Gola, G. Settimo, and S. Capolongo, "Indoor Air Quality in Inpatient Environments: A Systematic Review on Factors that Influence Chemical Pollution in Inpatient Wards," *Journal of Healthcare Engineering,* vol. 2019, p. 8358306, 2019/02/27 2019, doi: 10.1155/2019/8358306.

[6] M. M. Abdel-Salam, "Relationship between residential indoor air quality and socioeconomic factors in two urban areas in Alexandria, Egypt," *Building and Environment,* vol. 207, p. 108425, 2022.

[7] R. Abdel Sater, M. Perona, and C. Chevallier, "The effectiveness of personalised versus generic information in changing behaviour: Evidence from an indoor air quality experiment," Center for Open Science, 2021.

[8] N. R. Kapoor, A. Kumar, A. Kumar, A. Kumar, and H. C. Arora, "Prediction of Indoor Air Quality Using Artificial Intelligence," *Machine Intelligence, Big Data Analytics, and IoT in Image Processing: Practical Applications,* pp. 447-469, 2023.

[9] P. Anand, D. Cheong, and C. Sekhar, "A review of occupancy-based building energy and IEQ controls and its future post-COVID," *Science of the Total Environment,* vol. 804, p. 150249, 2022.

[10] R. R. Kureshi, D. Thakker, B. K. Mishra, and J. Barnes, "From Raising Awareness to a Behavioural Change: A Case Study of Indoor Air Quality Improvement Using IoT and COM-B Model," *Sensors,* vol. 23, no. 7, p. 3613, 2023.

[11] N. A. Megahed and E. M. Ghoneim, "Indoor Air Quality: Rethinking rules of building design strategies in post-pandemic architecture," *Environmental research,* vol. 193, p. 110471, 2021.

[12] S. Kirschke *et al.*, "Citizen science projects in freshwater monitoring. From individual design to clusters?," *Journal of Environmental Management,* vol. 309, p. 114714, 2022.

[13] T. van Noordwijk *et al.*, "Creating positive environmental impact through citizen science," *The science of citizen science,* pp. 373-395, 2021.

[14] R. R. Kureshi, D. Thakker, B. K. Mishra, and R. John, "AQ-SCIENCE: Air Quality - Smart Cities with IoT-ENabled Citizen Engagement Approach," presented at the 11th International Conference on the Internet of Things, St.Gallen, Switzerland, 2021. [Online]. Available: https://doi.org/10.1145/3494322.3494354.

[15] F. D. Davis, "A technology acceptance model for empirically testing new end-user information systems: Theory and results," Massachusetts Institute of Technology, 1985.

[16] L. C. Tagliabue, F. R. Cecconi, S. Rinaldi, and A. L. C. Ciribini, "Data driven indoor air quality prediction in educational facilities based on IoT

network," *Energy and Buildings,* vol. 236, p. 110782, 2021.
[17] W. S. Ismaeel and A. G. Mohamed, "Indoor air quality for sustainable building renovation: A decision-support assessment system using structural equation modelling," *Building and Environment,* vol. 214, p. 108933, 2022.
[18] R. R. Kureshi *et al.*, "Data-Driven Techniques for Low-Cost Sensor Selection and Calibration for the Use Case of Air Quality Monitoring," *Sensors,* vol. 22, no. 3, p. 1093, 2022. [Online]. Available: https://www.mdpi.com/1424-8220/22/3/1093.
[19] H. Chojer, P. Branco, F. Martins, M. Alvim-Ferraz, and S. Sousa, "Can data reliability of low-cost sensor devices for indoor air particulate matter monitoring be improved?–An approach using machine learning," *Atmospheric Environment,* vol. 286, p. 119251, 2022.
[20] C. Wang *et al.*, "How indoor environmental quality affects occupants' cognitive functions: A systematic review," *Building and environment,* vol. 193, p. 107647, 2021.
[21] F. Ibrahim, E. Z. Samsudin, A. R. Ishak, and J. Sathasivam, "Hospital indoor air quality and its relationships with building design, building operation, and occupant-related factors: A mini-review," *Frontiers in public health,* vol. 10, p. 1067764, 2022.
[22] M. Marzouk and M. Atef, "Assessment of Indoor Air Quality in Academic Buildings Using IoT and Deep Learning," *Sustainability,* vol. 14, no. 12, p. 7015, 2022.
[23] L.-R. Jia, J. Han, X. Chen, Q.-Y. Li, C.-C. Lee, and Y.-H. Fung, "Interaction between thermal comfort, indoor air quality and ventilation energy consumption of educational buildings: A comprehensive review," *Buildings,* vol. 11, no. 12, p. 591, 2021.
[24] C. De Capua, G. Fulco, M. Lugarà, and F. Ruffa, "An Improvement Strategy for Indoor Air Quality Monitoring Systems," *Sensors,* vol. 23, no. 8, p. 3999, 2023.
[25] C. C. Vassella *et al.*, "From spontaneous to strategic natural window ventilation: Improving indoor air quality in Swiss schools," *International Journal of Hygiene and Environmental Health,* vol. 234, p. 113746, 2021.
[26] A. N. Nair, P. Anand, A. George, and N. Mondal, "A review of strategies and their effectiveness in reducing indoor airborne transmission and improving indoor air quality," *Environmental Research,* vol. 213, p. 113579, 2022/10/01/ 2022, doi: https://doi.org/10.1016/j.envres.2022.113579.
[27] C. Jackson, C. Østerlund, V. Maidel, K. Crowston, and G. Mugar, "Which way did they go? Newcomer movement through the Zooniverse," in *Proceedings of the 19th ACM conference on computer-supported cooperative work & social computing*, 2016, pp. 624-635.
[28] T. B. Phillips, H. L. Ballard, B. V. Lewenstein, and R. Bonney, "Engagement in science through citizen science: Moving beyond data collection," *Science Education,* vol. 103, no. 3, pp. 665-690, 2019.
[29] J. A. Robinson *et al.*, "User-Centred Design of a Final Results Report for Participants in Multi-Sensor Personal Air Pollution Exposure Monitoring Campaigns," *International Journal of Environmental Research and Public Health,* vol. 18, no. 23, p. 12544, 2021. [Online]. Available: https://www.mdpi.com/1660-4601/18/23/12544.
[30] Y. N. Golumbic, B. Fishbain, and A. Baram-Tsabari, "User centered design of a citizen science air-quality monitoring project," *International Journal of Science Education, Part B,* vol. 9, no. 3, pp. 195-213, 2019.
[31] B. J. Hubbell *et al.*, "Understanding social and behavioral drivers and impacts of air quality sensor use," *Science of The Total Environment,* vol. 621, pp. 886-894, 2018.
[32] H. Pritchard and J. Gabrys, "From citizen sensing to collective monitoring: Working through the perceptive and affective problematics of environmental pollution," *GeoHumanities,* vol. 2, no. 2, pp. 354-371, 2016.
[33] S. Mahajan, C.-H. Luo, D.-Y. Wu, and L.-J. Chen, "From Do-It-Yourself (DIY) to Do-It-Together (DIT): Reflections on designing a citizen-driven air quality monitoring framework in Taiwan," *Sustainable Cities and Society,* vol. 66, p. 102628, 2021.
[34] P. B. English *et al.*, "The Imperial County Community Air Monitoring Network: a model for community-based environmental monitoring for public health action," *Environmental health perspectives,* vol. 125, no. 7, p. 074501, 2017.
[35] C. Leonardi, A. Cappellotto, M. Caraviello, B. Lepri, and F. Antonelli, "SecondNose: an air quality mobile crowdsensing system," in *Proceedings of the 8th Nordic Conference on Human-Computer Interaction: Fun, Fast, Foundational*, 2014, pp. 1051-1054.
[36] S. Y. Yousafzai, G. R. Foxall, and J. G. Pallister, "Technology acceptance: a meta‐analysis of the TAM: Part 1," *Journal of Modelling in Management,* vol. 2, no. 3, pp. 251-280, 2007, doi: 10.1108/17465660710834453.
[37] T. Zhou, "An empirical examination of initial trust in mobile banking," *Internet Research,* vol. 21, no. 5, pp. 527-540, 2011.
[38] T. Laukkanen, "Consumer adoption versus rejection decisions in seemingly similar service innovations: The case of the Internet and mobile banking," *Journal of Business Research,* vol. 69, no. 7, pp. 2432-2439, 2016/07/01/ 2016, doi: https://doi.org/10.1016/j.jbusres.2016.01.013.
[39] S. S. Kim, N. K. Malhotra, and S. Narasimhan, "Research note—two competing perspectives on automatic use: A theoretical and empirical comparison," *Information systems research,* vol. 16, no. 4, pp. 418-432, 2005.
[40] H. Taherdoost, "A review of technology acceptance and adoption models and theories," *Procedia Manufacturing,* vol. 22, pp. 960-967, 2018/01/01/

2018, doi: https://doi.org/10.1016/j.promfg.2018.03.137.

[41] J. I. Campbell *et al.*, "The Technology Acceptance Model for Resource-Limited Settings (TAM-RLS): A Novel Framework for Mobile Health Interventions Targeted to Low-Literacy End-Users in Resource-Limited Settings," *AIDS and Behavior,* vol. 21, no. 11, pp. 3129-3140, 2017/11/01 2017, doi: 10.1007/s10461-017-1765-y.

[42] Z. Hu, S. Ding, S. Li, L. Chen, and S. Yang, "Adoption intention of fintech services for bank users: An empirical examination with an extended technology acceptance model," *Symmetry,* vol. 11, no. 3, p. 340, 2019.

[43] V. Chooprayoon and C. Fung, "TECTAM: An Approach to Study Technology Acceptance Model (TAM) in Gaining Knowledge on the Adoption and Use of E-Commerce/E-Business Technology among Small and Medium Enterprises in Thailand," 2010.

[44] A. Granić, "Educational Technology Adoption: A systematic review," *Education and Information Technologies,* vol. 27, no. 7, pp. 9725-9744, 2022/08/01 2022, doi: 10.1007/s10639-022-10951-7.

[45] V. Venkatesh and F. D. Davis, "A Theoretical Extension of the Technology Acceptance Model: Four Longitudinal Field Studies," *Management Science,* vol. 46, no. 2, pp. 186-204, 2000. [Online]. Available: http://www.jstor.org/stable/2634758.

[46] A. M. Al-Rahmi *et al.*, "The Influence of Information System Success and Technology Acceptance Model on Social Media Factors in Education," *Sustainability,* vol. 13, no. 14, p. 7770, 2021. [Online]. Available: https://www.mdpi.com/2071-1050/13/14/7770.

[47] S. R. Simon *et al.*, "Physicians and electronic health records: a statewide survey," *Archives of internal medicine,* vol. 167, no. 5, pp. 507-512, 2007.

[48] S. A. Kamal, M. Shafiq, and P. Kakria, "Investigating acceptance of telemedicine services through an extended technology acceptance model (TAM)," *Technology in Society,* vol. 60, p. 101212, 2020.

[49] P. R. Palos-Sanchez, J. R. Saura, M. A. Rios Martin, and M. Aguayo-Camacho, "Toward a better understanding of the intention to use mHealth apps: exploratory study," *JMIR mHealth and uHealth,* vol. 9, no. 9, p. e27021, 2021.

[50] C. M. Mao and S. R. Hovick, "Adding affordances and communication efficacy to the technology acceptance model to study the messaging features of online patient portals among young adults," *Health Communication,* vol. 37, no. 3, pp. 307-315, 2022.

[51] T.-H. Tsai, W.-Y. Lin, Y.-S. Chang, P.-C. Chang, and M.-Y. Lee, "Technology anxiety and resistance to change behavioral study of a wearable cardiac warming system using an extended TAM for older adults," *PloS one,* vol. 15, no. 1, p. e0227270, 2020.

[52] P. Esmaeilzadeh, "Use of AI-based tools for healthcare purposes: a survey study from consumers' perspectives," *BMC medical informatics and decision making,* vol. 20, no. 1, pp. 1-19, 2020.

[53] A. Steinemann, P. Wargocki, and B. Rismanchi, "Ten questions concerning green buildings and indoor air quality," *Building and Environment,* vol. 112, pp. 351-358, 2017.

[54] Y. Geng, W. Ji, Z. Wang, B. Lin, and Y. Zhu, "A review of operating performance in green buildings: Energy use, indoor environmental quality and occupant satisfaction," *Energy and Buildings,* vol. 183, pp. 500-514, 2019.

[55] F. D. Davis and V. Venkatesh, "A critical assessment of potential measurement biases in the technology acceptance model: three experiments," *International journal of human-computer studies,* vol. 45, no. 1, pp. 19-45, 1996.